\documentclass{article}
\def\abstract#1{\vskip 7mm 
        \begin{center}{\large Abstract}\par \smallskip
                \begin{minipage}[c]{12cm}
                        \small #1
                \end{minipage}
        \end{center}
}
\def\title#1{\begin{center}{\Large\bf #1}\end{center}}
\def\author#1{\vskip 5mm \begin{center}{#1}\end{center}}
\def\address#1{\begin{center}{\it #1}\end{center}}
\makeatletter
\@ifundefined{lesssim}{}{}
\@ifundefined{gtrsim}{}{}
\def\vereq#1#2{\lower3pt\vbox{\baselineskip1.5pt \lineskip1.5pt
\ialign{$\m@th#1\hfill##\hfil$\crcr#2\crcr\sim\crcr}}}
\makeatother

\begin{document}

\title{%
  Regular second order perturbations of extreme mass ratio black hole binaries
  \smallskip \\
  {\large }
}
\author{%
  Hiroyuki Nakano\footnote{E-mail:nakano@phys.utb.edu}
  and
  Carlos O. Lousto\footnote{E-mail:lousto@phys.utb.edu}
}
\address{%
  Department of Physics and Astronomy, 
  and Center for Gravitational Wave Astronomy, \\
The University of Texas at Brownsville, Brownsville, Texas 78520, USA
\\
and\\%}
%\address{%
Center for Computational Relativity and Gravitation,
School of Mathematical Sciences, \\
Rochester Institute of Technology, 78 Lomb Memorial Drive, Rochester,
 New York 14623, USA
}
\abstract{
  We report on the first results of self-consistent second 
  order metric perturbations produced 
  by a point particle moving in the Schwarzschild 
  spacetime. The second order waveforms satisfy a wave equation with an 
  effective source term build up from products of first order metric 
perturbations 
  and its derivatives. We have explicitly 
regularized this source term at the particle
  location as well as at the horizon and at spatial infinity.
}

%%%%%%%%%%%%%%%%%%%%%%%%%%%%%%%%%%%%%%%%%%%%%%%%%%%%%%%%%%%%%%%%%%%%%%
\section{Introduction}
%%%%%%%%%%%%%%%%%%%%%%%%%%%%%%%%%%%%%%%%%%%%%%%%%%%%%%%%%%%%%%%%%%%%%%

The observation of gravitational waves opens 
 a new window onto our universe and
we also expect that the observation of gravitational waves 
will provides a direct experimental test of general relativity. 

The space mission LISA will primarily detect gravitational waves from 
inspiraling solar-mass compact objects 
captured by a supermassive black hole residing in the core of active
galaxies.
For these Extreme Mass Ratio Inspirals (EMRI) 
we use the black hole perturbation approach, 
where the compact object is approximated by a point particle 
orbiting a massive black hole. 
There are two nontrivial problems to consider in this approach: 
The self-force problem and the second order gravitational 
perturbations problem. 
Due to the self-force 
the orbit of the particle deviates from the background geodesic, 
i.e. the spacetime is perturbed by the particle itself. 
It is essential to take this deviation into account 
in order to predict the orbital evolution to the required order. 
For the headon configuration studied in this paper this was
achieved in~\cite{Barack:2002ku}.
The gravitational self-force is, however, not easily obtainable
for more general trajectories. 

We require the second perturbative order calculations to derive 
the precise gravitational waveforms to be used as templates for 
gravitational wave data analysis. 
In general, this computation has to be done by numerical integration. 
Hence, it is important to derive a well-behaved second order effective source.
In this paper, we will focus on this later problem.

%%%%%%%%%%%%%%%%%%%%%%%%%%%%%%%%%%%%%%%%%%%%%%%%%%%%%%%%%%%%%%%%%%%%%%
\section{Second order metric perturbations} 
%%%%%%%%%%%%%%%%%%%%%%%%%%%%%%%%%%%%%%%%%%%%%%%%%%%%%%%%%%%%%%%%%%%%%%

We consider second order metric perturbations (MP), 
%\begin{eqnarray}
$\tilde g_{\mu\nu}=g_{\mu\nu}+h_{\mu\nu}^{(1)}+h_{\mu\nu}^{(2)} \,,$
%\nonumber 
%\end{eqnarray}
with expansion parameter $\mu/M$ corresponding to the mass ratio of the holes
and where $g_{\mu\nu}$ is the Schwarzschild metric. 
The Hilbert-Einstein tensor and the stress-energy tensor up to 
the second perturbative order is given by 
\begin{eqnarray}
G_{\mu\nu}[\tilde g_{\mu\nu}] = 
G_{\mu\nu}^{(1)}[h^{(1)}]+G_{\mu\nu}^{(1)}[h^{(2)}]+G_{\mu\nu}^{(2)}[h^{(1)},h^{(1)}] 
\,, \quad 
T_{\mu\nu} =
T_{\mu\nu}^{(1)}+T_{\mu\nu}^{(2,SF)}+T_{\mu\nu}^{(2,h)} 
\label{eq:T-exp} \,,
\end{eqnarray}
where
\begin{eqnarray}
G_{\mu\nu}^{(1)}[h] &=&
-\frac{1}{2}h_{\mu\nu;\alpha}{}^{;\alpha}+h_{\alpha(\mu;\nu)}{}^{;\alpha}
-R_{\alpha\mu\beta\nu}h^{\alpha\beta}
-\frac{1}{2}h_{;\mu\nu} -\frac{1}{2} g_{\mu\nu}
(h_{\lambda\alpha}{}^{;\alpha\lambda}-h_{;\lambda}{}^{;\lambda}) \,,
\nonumber 
\\ 
G_{\mu\nu}^{(2)}[h^{(1)},h^{(1)}] &=& 
R_{\mu\nu}^{(2)}[h^{(1)},h^{(1)}] - \frac{1}{2}g_{\mu\nu}R^{(2)}[h^{(1)},h^{(1)}] 
\,; \nonumber \\
R_{\mu\nu}^{(2)}[h^{(1)},h^{(1)}] 
&=& \frac{1}{4}h^{(1)}_{\alpha\beta;\mu}h^{(1)}{}^{\alpha\beta}{}_{;\nu}
+\frac{1}{2}h^{(1)}{}^{\alpha\beta}(h^{(1)}_{\alpha\beta;\mu\nu}
+h^{(1)}_{\mu\nu;\alpha\beta}
-2h^{(1)}_{\alpha(\mu;\nu)\beta}) \nonumber \\ 
&& 
-\frac{1}{2}(h^{(1)}{}^{\alpha\beta}{}_{;\beta}-\frac{1}{2}h^{(1)}{}^{;\alpha})
(2h^{(1)}_{\alpha(\mu;\nu)}-h^{(1)}_{\mu\nu;\alpha})
+\frac{1}{2}h^{(1)}_{\mu\alpha;\beta}h^{(1)}_{\nu}{}^{\alpha;\beta} 
-\frac{1}{2}h^{(1)}_{\mu\alpha;\beta}h^{(1)}_{\nu}{}^{\beta;\alpha} \,.
\nonumber 
\end{eqnarray}
On the other hand, the stress-energy tensor includes 
three parts, the first order stress-energy tensor  $T_{\mu\nu}^{(1)}$ 
which is the one of a point particle moving along a background geodesic; 
\begin{eqnarray}
T^{(1)}{}^{\mu\nu} &=& \mu \int^{+\infty}_{-\infty} \delta^{(4)}(x-z(\tau))
{dz^{\mu} \over d\tau}{dz^{\nu} \over d\tau}d\tau \,, 
\label{eq:pp}
\end{eqnarray}
where $z^{\mu}=\{T(\tau),R(\tau),\Theta(\tau),\Phi(\tau)\}$ 
for the particle orbit, 
the deviation from the geodesic 
by the self-force $T_{\mu\nu}^{(2,SF)}$ 
(See Ref.~ \cite{Barack:2002ku}), 
and finally $T_{\mu\nu}^{(2,h)}$, which is purely affected by 
the first order MP, 
\begin{eqnarray}
T_{\mu\nu}^{(2,h)} = - \frac{1}{2} \mu \int^{+\infty}_{-\infty} 
h^{(1)} \,\delta^{(4)}(x-z(\tau))
{dz^{\mu} \over d\tau}{dz^{\nu} \over d\tau}d\tau \,,
\end{eqnarray}
where we have used 
the determinant $\tilde g=g(1+h^{(1)})$ up to the first perturbative order.

%%%%%%%%%%%%%%%%%%%%%%%%%%%%%%%%%%%%%%%%%%%%%%%%%%%%%%%%%%%%%%%%%%%%%%
\section{First order metric perturbations in the RW gauge} \label{sec:1st}
%%%%%%%%%%%%%%%%%%%%%%%%%%%%%%%%%%%%%%%%%%%%%%%%%%%%%%%%%%%%%%%%%%%%%%

Before considering the second order, 
it is necessary to discuss the first order MP, 
i.e.,  the Regge-Wheeler-Zerilli formalism~\cite{Regge:1957td,Zerilli:wd}. 
The basic formalism has been given in Zerilli's paper~\cite{Zerilli:wd}, 
and it has been summarized in the time domain in~\cite{Lousto:2005ip,Lousto:2005xu}.
In the following, equation numbers (Z:1), (L1:1) and (L2:1) for instance, 
denote the equation (1) in~\cite{Zerilli:wd}, \cite{Lousto:2005ip} 
and~\cite{Lousto:2005xu}, respectively. 

For the first order Hilbert-Einstein equation, 
$G_{\mu\nu}^{(1)}[h^{(1)}] = 8 \, \pi \,T_{\mu\nu}^{(1)}$, 
we expand $h_{\mu\nu}^{(1)}$ 
and $T_{\mu\nu}^{(1)}$ in ten tensor harmonics components. 
We then obtain the linearized field equations for each harmonic mode. 
For the even part, which has the even parity behavior, $(-1)^{\ell}$, 
we have seven equations. 
We impose the Regge-Wheeler gauge conditions (RW), the vanishing of
some coefficients of the first order MP: $h_{0}^{(e)}=h_{1}^{(e)}=G=0$.

We introduce the following wave-function for the even parity modes, 
\begin{eqnarray}
\psi^{\rm even}_{\ell m}(t,r)&=&
\frac{2\,r}{\ell(\ell+1)} \left[
K_{\ell m}^{\rm RW} (t,r)
+2\,{\frac { ( r-2\,M )  }
{ ( r{\ell}^{2}+r\ell-2\,r+6\,M ) }}
\left(H_{2\,\ell m}^{\rm RW} ( t,r )
-r\,
{\frac {\partial }{\partial r}}K_{\ell m}^{\rm RW}(t,r)\right)
\right]
\,,
\label{eq:defpsi}
\end{eqnarray}
where the suffix ${\rm RW}$ stands for the RW gauge. 
This function $\psi^{\rm even}_{\ell m}$ obeys the Zerilli equation, 
\begin{eqnarray}
\hat{\cal Z}_\ell^{\rm even}  \,\psi^{\rm even}_{\ell m}(t,r) &=& S^{\rm even}_{\ell m}(t,r) \,; 
\quad 
\hat{\cal Z}_\ell^{\rm even} = -\frac{\partial^2}{{\partial t}^2} 
+ \frac{\partial^2}{{\partial r^*}^2}-V_\ell^{\rm even}(r) \,,
\label{eq:zerilli}
\end{eqnarray}
where $r^*=r+2M \log (r/2M-1)$, the potential $V_{\ell}^{{\rm even}}$ 
and the source $S_{\ell m}^{\rm even}$ are given in Eqs.~(L1:1-2) 
and (L1:A.3). The reconstruction of the MP 
under the RW gauge have been expressed in Eqs.~(L2:B.9-12). 

In the following, we consider a particle falling radially 
into a Schwarzschild black hole as the first order source. 
The equation of motion of the test particle is given by 
\begin{eqnarray}
\left(\frac{dR}{dt}\right)^2 &=& -\left(1-\frac{2M}{R}\right)^3 \frac{1}{E^2}
+ \left(1-\frac{2M}{R}\right)^2 \,; \quad 
E = \left(1-\frac{2M}{R} \right)\,\frac{dT(\tau)}{d\tau} \,,
\end{eqnarray}
where $E$ and $R$ are the energy and the location of the particle, respectively. 
The non-vanishing tensor harmonics coefficients of the energy-momentum tensor 
are $A_{\ell m}$, $A_{\ell m}^{(0)}$ and $A_{\ell m}^{(1)}$. 
Because of the symmetry of the problem we have only to consider even parity modes, 
i.e., described by the Zerilli equation. 

In the head on collision case, the MP in the RW gauge 
are $C^0$ (continuous across the particle). One can see this as follows.
First, from Eqs.~(Z:C7d) and (Z:C7e), we find 
that $H_{2\,\ell m}(=H_{0\,\ell m})$ and $K_{\ell m}$ have 
the same differential behavior. 
Then, we note that $\partial_r K_{\ell m} \sim \theta(r-R(t))$, 
because the left hand side of (\ref{eq:defpsi}) behaves 
as a step function near the particle. 
This means that $K_{\ell m}$ (and also $H_{2\,\ell m}$) is $C^0$. 
Using this, $\partial_r H_{1\,\ell m} \sim \theta(r-R(t))$ 
is derived from Eq.~(Z:C7e), i.e., $H_{1\,\ell m}$ is $C^0$.
(See Ref.~\cite{Lousto:1999za}.) 

Using the above fact, we can take up second derivatives of 
the function $\psi^{\rm even}_{\ell m}$ 
with respect to $t$ and $r$. 
These quantities allow us to calculate 
the coefficients of the $\delta$-terms in the second order source.

Next, we consider the $\ell=0$ perturbations ($\ell=1$ modes can be 
completely eliminated in the center of mass coordinate system). 
In the RW gauge there are four non-vanishing coefficients of the MP, 
$H_{0\,00}$, $H_{1\,00}$, $H_{2\,00}$ and 
$K_{00}$. The gauge transformation has two extra degrees of freedom. 
We could choose the gauge so that 
$H_{1\,00}=K_{00}=0$, i.e., the Zerilli gauge, 
but it is difficult to treat the second order source, 
since the MP are not $C^0$. 

We instead consider here a new (singular) gauge transformation, 
chosen to make the metric perturbations $C^0$ 
and to obtain an appropriate second order source behavior. 
To derive this gauge transformation, we have also considered 
a regularization of the second order source at $r=\infty$ 
and the horizon at the same time~\footnote{
We could regularize and fix the gauge for the second order source. 
One could also proceed by first fixing the gauge.}.
Note that we succeeded in choosing the gauge 
such that all of the above MP 
behave as $C^0$ at the location of the particle and vanish 
at $r=\infty$ and $r=2M$.

%%%%%%%%%%%
\section{Second order Zerilli equation}
%%%%%%%%%%%

Since the first order MP 
contains only even parity modes, we can discuss the second order MP 
for the even parity modes only, i.e. in terms of the Zerilli function, 
\begin{eqnarray}
\chi_{20}^{\rm Z}(t,r)
&=&
\frac{1}{2\,r+3\,M}
\left({r}^2{\frac {\partial }{\partial t}}{\cal K}_{20} ( t,r )
-( r-2\,M ) {\cal H}_{1\,20} ( t,r ) \right) \,.
\end{eqnarray}
Here, we have considered the contribution 
from the $\ell=0$ and $2$ modes of the first order
to the $\ell=2$ mode of the second order since this gives the leading 
contribution to gravitational radiation. We also choose 
the RW gauge to second order. 
This Zerilli function satisfies the equation, 
$\hat{\cal Z}_2^{\rm even} \chi_{20}^{\rm Z}(t,r) 
= {\cal S}^{\rm{Z}}_{20}(t,r)$ with 
\begin{eqnarray}
{\cal S}^{\rm{Z}}_{20}(t,r) &=& 
 {\frac {8\,\pi \,\sqrt {3} \left( r-2\,M \right) ^{2}}
{3(2\,r+3\,M)}}\,{\frac {\partial }{\partial t}}{\cal B}_{20} ( t,r ) 
+ {\frac { 8\,\pi \left( r-2\,M \right) ^{2}}{2\,r+3\,M}}
\,{\frac {\partial }{\partial t}} {\cal A}_{20} ( t,r ) 
- \frac{8\,\sqrt {3}\, \pi \left( r-2\,M \right) }{3}
{\frac {\partial }{\partial t}}{\cal F}_{20} ( t,r ) 
\nonumber \\ && 
- {\frac {4\,\sqrt {2}\,i\pi  \left( r-2\,M \right) ^{2} }
{2\,r+3\,M}}{\frac {\partial }{\partial r}}{\cal A}^{(1)}_{20} ( t,r ) 
- {\frac {8\,\sqrt {2}\,i\pi  \left( r-2\,M \right)  \left( 5\,r-3\,M \right) M}
{r \left( 2\,r+3\,M \right) ^{2}}}{\cal A}^{(1)}_{20} ( t,r ) 
\nonumber \\ && 
- {\frac {8\,\sqrt {3} \,i\pi \left( r-2\,M \right) ^{2}}{3(2\,r+3\,M)}}
{\frac {\partial }{\partial r}}{\cal B}^{(0)}_{20} ( t,r ) 
+ {\frac {32\,\sqrt {3}\,i\pi \,
\left( 3\,{M}^{2}+{r}^{2} \right)  \left( r-2\,M \right) }
{3\,r \left( 2\,r+3\,M \right) ^{2}}} {\cal B}^{(0)}_{20} ( t,r ) 
\,.
\label{eq:2ndS22}
\end{eqnarray}
The functions ${\cal B}_{20}$ etc. are derived 
from the second order quantities, $G_{\mu\nu}^{(2)}[h^{(1)},h^{(1)}]$, 
$T_{\mu\nu}^{(2,h)}$ (and $T_{\mu\nu}^{(2,SF)}$) 
by the same tensor harmonics expansion as for the first order. 

The delta function $\delta^{(4)}(x-z(\tau))$
in $T_{\mu\nu}^{(2,h)}$ includes 
%an angular dependence 
$\delta^{(2)}(\Omega-\Omega(\tau))
=\sum_{\ell m}Y_{\ell m}(\Omega)Y_{\ell m}^*(\Omega(\tau))$. 
We have considered only the contribution from the $\ell=0$ and $2$ modes 
of the first order perturbations. Consistently, we use only the three components, 
$h^{(1)}_{(\ell=2)}Y_{2m}(\Omega)Y_{2m}^*(\Omega(\tau))$, 
$h^{(1)}_{(\ell=2)}Y_{0m}(\Omega)Y_{0m}^*(\Omega(\tau))$ and 
$h^{(1)}_{(\ell=0)}Y_{2m}(\Omega)Y_{2m}^*(\Omega(\tau))$. 

We may wander if there is any $\delta^2$-term in the second order source.
The answer is ``No''. This is because 
in the case of the head on collision, 
the MP under the RW gauge are $C^0$: 
$R_{\mu\nu}^{(2)}[h^{(1)},h^{(1)}]$ includes 
second derivatives and 
we need one more derivative to construct ${\cal S}^{\rm{Z}}_{20}$.
$(h^{(1)})^2$ is $C^0 \times C^0$ and 
its third derivative yields $C^0 \times \delta'$ and 
$\theta \times \delta$ as the most singular terms. 
On the other hand, $T_{\mu\nu}^{(2,h)}$ includes 
only $C^0 \times \delta$ terms. 

From Eq.~(\ref{eq:2ndS22}), we obtain the second order source as 
\begin{eqnarray}
{\cal S}^{\rm{Z}}_{20}(t,r) &=& {}^{(2,2)}{\cal S}^{\rm{Z}}_{20}(t,r)
+{}^{(0,2)}{\cal S}^{\rm{Z}}_{20}(t,r) \,, 
\end{eqnarray}
where ${}^{(2,2)}{\cal S}^{\rm{Z}}_{20}$ and ${}^{(0,2)}{\cal S}^{\rm{Z}}_{20}$ 
are the contribution from $(\ell=2)\cdot(\ell=2)$ and 
$(\ell=0)\cdot(\ell=2)$, respectively. 

Note that while the above source term is locally well behaved
near the particle location, some terms diverge as $r\to\infty$.
So, we need to consider some regularization for the asymptotic behavior \cite{Gleiser:1998rw}. 

In order to obtain a well behaved source for large values of $r$, 
we define a renormalized Zerilli function by 
\begin{eqnarray}
{\tilde \chi}_{20}^{\rm Z}(t,r) &=& \chi_{20}^{\rm Z}(t,r) 
- \chi_{20}^{{\rm reg},(2,2)} - \chi_{20}^{{\rm reg},(0,2)} \,; 
\\ 
\chi_{20}^{{\rm reg},(2,2)} &=& 
\frac{\sqrt{5}}{7\,\sqrt{\pi}}
{\frac {{r}^{2}}{2\,r+3\,M}} 
\left( {\frac {\partial }{\partial t}}K_{20}^{\rm RW} ( t,r )  \right) 
K_{20}^{\rm RW} ( t,r ) \,, 
\nonumber \\ 
\chi_{20}^{{\rm reg},(0,2)} &=& 
- \frac{1}{32\sqrt {\pi }}\,\frac{r \left(r-2\,M \right)}{ 2\,r+3\,M} 
\Biggl( 
 \left( 1-\frac{M}{r}\ln  \left( {\frac {r}{2M}} \right) \right) 
H_{2\,00}^{\rm RZ}(t,r) {\frac {\partial }{\partial r}}H_{2\,20}^{\rm RW} ( t,r ) 
\nonumber \\ && \,
+6\, \left( 1-5\,{\frac {M}{r}} \right) H_{0\,00}^{\rm RZ}(t,r)
{\frac {\partial }{\partial r}}H_{2\,20}^{\rm RW} ( t,r ) 
+4\, \left( 1+3\,\frac{M}{r}\ln  \left({\frac {r}{2M}} \right)  \right) 
H_{2\,00}^{\rm RZ}(t,r) {\frac {\partial }{\partial t}}H_{2\,20}^{\rm RW} ( t,r ) 
 \nonumber \\ && \,
+5\, \left( 1-4\,{\frac {M}{r}} \right) H_{0\,00}^{\rm RZ}(t,r)
{\frac {\partial }{\partial t}}H_{2\,20}^{\rm RW} ( t,r ) 
-18\,\frac{1}{r}\,\ln  \left( {\frac {r}{2M}} \right) H_{0\,00}^{\rm RZ}(t,r)
H_{2\,20}^{\rm RW} ( t,r )  \Biggr) \,, 
\nonumber 
\end{eqnarray}
where the suffix ${\rm RZ}$ means some gauge choice as discussed in Sec.~\ref{sec:1st}. 
Finally, 
The best suited equation to solve numerically for ${\tilde \chi}_{20}^{\rm Z}$ is then
\begin{eqnarray}
\hat{\cal Z}_2^{\rm even} 
{\tilde \chi}_{20}^{\rm Z}(t,r) &=& {\cal S}^{\rm{Z},reg}_{20}(t,r) \,;
\\
{\cal S}^{\rm{Z},reg}_{20}(t,r) &=& 
\left({}^{(2,2)}{\cal S}^{\rm{Z}}_{20}(t,r) 
- \hat{\cal Z}_2^{\rm even} \chi_{20}^{{\rm reg},(2,2)}(t,r)\right)
+\left({}^{(0,2)}{\cal S}^{\rm{Z}}_{20}(t,r)
- \hat{\cal Z}_2^{\rm even} \chi_{20}^{{\rm reg},(0,2)}(t,r)\right) \,. 
\nonumber 
\end{eqnarray}

%%%%%%%%%%%%%%%%%%%%%%%%%%%%%%%%%%%%%%%%%%%%%%%%%%%%%%%%%%%%%%%%%%%%%%
\section{Discussion} 
%%%%%%%%%%%%%%%%%%%%%%%%%%%%%%%%%%%%%%%%%%%%%%%%%%%%%%%%%%%%%%%%%%%%%%

In this paper, we obtained the regularized second order effective source 
of the Zerilli equation in the case of a particle falling radially 
into a Schwarzschild black hole. 
Using this source, 
we are able to compute the second order contribution to gravitational radiation
by numerical integration. 

To be fully second order consistent we have to include the term $T_{\mu\nu}^{(2,SF)}$ which 
is derived from the self-force on a particle. 
The self-force for a headon collision has been calculated 
in~\cite{Barack:2002ku}, 
and in a circular orbit around a Schwarzschild black hole in~\cite{Nakano:2003he}, 
but have not been obtained in the general case yet. 

To prove that 
there is no $\delta^2$ term in the second order source, 
we have used the fact that 
the first order MP in the RW gauge is $C^0$. 
In the general orbit cases (including circular orbits), 
the first order MP is not $C^0$ in the RW gauge, but 
it is $C^0$ in the Lorenz gauge \cite{Barack:2005nr}.
This gauge choice favors the study of the second order perturbations
for generic orbits.

%===========================%
\subsection*{Acknowledgments}
%===========================%

We would like to thank K.~Nakamura and T.~Tanaka for useful discussions.

\end{document}